\documentclass[titlepage,12pt]{article}
\usepackage{amsmath}
\usepackage{amssymb}
\usepackage{amsfonts}
\usepackage{geometry}
\usepackage{mathrsfs}
\usepackage[onehalfspacing]{setspace}
\usepackage{rotating}

\newtheorem{theorem}{Theorem}

\input{tcilatex}
\renewcommand{\mathcal}{\mathscr}

\begin{document}

\title{Spatial kriging for replicated temporal point processes}
\author{Daniel Gervini \\
Department of Mathematical Sciences\\
University of Wisconsin--Milwaukee}
\maketitle

\begin{abstract}
This paper presents a kriging method for spatial prediction of temporal
intensity functions, for situations where a temporal point process is
observed at different spatial locations. Assuming that several replications
of the processes are available at the spatial sites, this method avoids
assumptions like isotropy, which are not valid in many applications. As part
of the derivations, new nonparametric estimators for the mean and covariance
functions of temporal point processes are introduced, and their properties
are studied theoretically and by simulation. The method is applied to the
analysis of bike demand patterns in the Divvy bicycle sharing system of the
city of Chicago.

\end{abstract}

\section{Introduction\label{sec:Introd}}

In recent years, bicycle sharing systems have become increasingly common in
large cities around the world (Shaheen et al., 2010). In these systems,
customers can pick up and return bicycles at automated bike stations
distributed within a city. Although the stations are automated, maintenance
of the system requires active human management. The main issue is flow
imbalance: for example, most trips during the morning commute hours tend to
flow from the outer neighborhoods towards the city downtown, so neighborhood
stations would quickly run out of bikes and downtown stations would quickly
fill up if they were not rebalanced by trucking bikes from full stations to
empty stations (Nair and Miller-Hooks, 2011). These operations, to run
efficiently, require knowledge of spatio-temporal patterns of bike demand in
the city. Other important aspects of system management include decisions to
open, close, or relocate bike stations. To open a new bike station, the
managers have to be able to forecast the patterns of bike demand at the
intended new location. This is not simple, because these patterns may vary
greatly within short distances (Gervini and Khanal, 2019).

From a mathematical point of view, bike demand can be modelled by
spatio-temporal point processes (Gervini and Khanal, 2019). This can be done
in several ways, depending on the researchers' goals. In this paper, we
model bike check-out times at each station as a replicated temporal point
process, where each day of the year is a replication of the process.
Therefore, for a bike station located at spatial location $\mathbf{s}_{j}$,
we will have $n$ intensity functions $\{\lambda _{i}^{j}(t)\}_{i=1}^{n}$,
one for each day. The availability of replications allows us to estimate the
spatial mean and covariance of these intensity functions nonparametrically,
without resorting to assumptions like isotropy which are not valid in this
context (Gervini and Khanal, 2019). These estimators, in turn, can be used
to predict the intensity functions $\lambda _{i}^{0}(t)$'s at a new spatial
site $\mathbf{s}_{0}$.

In the literature, the usual approach to spatial prediction is kriging
(Cressie, 1993). Kriging has been extended to functional data contexts
(Giraldo et al., 2010, 2011; Menafoglio et al., 2013). These methods could
be applied in our context if the intensity functions at each site were
estimable by smoothing, but this is not always possible because of the low
daily counts at some bike stations. More importantly, these methods were
developed for situations where only one observation per site is available,
which makes assumptions like isotropy unavoidable.

In this paper we present a kriging method for spatial prediction of temporal
intensity functions that can be applied in anisotropic situations, when
several replications of the processes are available at the spatial sites. As
part of this method we introduce new nonparametric estimators for the mean
and covariance functions of temporal point processes and study their
properties.

\section{Models and methods}

\subsection{Poisson point processes\label{sec:Point_processes}}

A temporal point process $X$ is a random countable set in $\mathcal{S}%
\subseteq (0,\infty )$ (M\o ller and Waagepetersen, 2004, ch.~2). The
process is locally finite if $\#(X\cap B)<\infty $ with probability one for
any bounded $B\subseteq \mathcal{S}$, where $\#$ denotes the cardinality of
a set. In that case we define the count function $N(B)=\#(X\cap B)$ for each
bounded $B\subseteq \mathcal{S}$. In particular, we define $N(t)=\#(X\cap
(0,t])$. A Poisson process is a locally finite process for which there
exists a locally integrable function $\lambda :\mathcal{S}\rightarrow
\lbrack 0,\infty )$, called the intensity function, such that \emph{(i)} $%
N(B)$ has a Poisson distribution with rate $\int_{B}\lambda (t)~dt$, and 
\emph{(ii)} for disjoint sets $B_{1},\ldots ,B_{k}$ in $\mathcal{S}$ the
random variables $N(B_{1}),\ldots ,N(B_{k})$ are independent. A consequence
of \emph{(i)} and \emph{(ii)} is that the conditional distribution of the
points in $X\cap B$ given $N(B)=m$ is the distribution of $m$ independent
and identically distributed observations with density $\lambda
(t)/\int_{B}\lambda $. In this paper we will mostly consider temporal
processes defined on a common bounded interval $\mathcal{S}=[a,b]$, for
example $\mathcal{S}=[0,24]$ for daily processes.

For replicated point processes, a single intensity function $\lambda $
rarely provides an adequate fit for all replications; it is more reasonable
to assume that $\lambda $ itself is the realization of a random process $%
\Lambda $ and thus changes from replication to replication. Such compound
processes are called doubly stochastic or Cox processes (M\o ller and
Waagepetersen, 2004, ch.~5). A doubly stochastic Poisson process is a pair $%
(X,\Lambda )$ where $X|\Lambda =\lambda $ is a Poisson process with
intensity function $\lambda $, and $\Lambda $ is a random function that
takes values on the space $\mathcal{F}$ of non-negative locally integrable
functions on $\mathcal{S}$. Thus the $n$ daily replications of the process
can be modeled as $n$ independent and identically distributed pairs $%
(X_{1},\Lambda _{1}),\ldots ,(X_{n},\Lambda _{n})$. The latent process $%
\Lambda $ is not directly observable; only $X$ is observed.

In our applications we observe temporal processes at $d$ different spatial
locations. This can be modeled as a multivariate doubly stochastic process $(%
\mathbf{X},\mathbf{\Lambda })$ with $\mathbf{X}=(X^{1},\ldots ,X^{d})$ and $%
\mathbf{\Lambda }=(\Lambda ^{1},\ldots ,\Lambda ^{d})$, where the $X^{j}$s
are conditionally independent given $\mathbf{\Lambda }=\mathbf{\lambda }$.
The dependencies among the $X^{j}$s are then determined by the dependencies
among the $\Lambda ^{j}$s. Since each $\Lambda ^{j}$ is associated with a
specific spatial location $\mathbf{s}_{j}$, we will make this explicit in
the notation by writing $\Lambda ^{j}(t)=\Lambda (t,\mathbf{s}_{j})$, but $%
\Lambda (t,\mathbf{s})$ is not a joint spatio-temporal intensity function,
it is only a temporal intensity function in $t$ for each $\mathbf{s}$.

\subsection{Spatial kriging\label{sec:Kriging}}

The unbiased kriging predictor of $\Lambda (t,\mathbf{s}_{0})$, the
intensity process at a new spatial location $\mathbf{s}_{0}$, based on $%
\Lambda (t,\mathbf{s}_{1}),\ldots ,\Lambda (t,\mathbf{s}_{d})$ is 
\begin{equation*}
\Lambda ^{\ast }(t,\mathbf{s}_{0})=\sum_{j=1}^{d}c_{j}^{\ast }\Lambda (t,%
\mathbf{s}_{j}),
\end{equation*}%
where $\mathbf{c}^{\ast }=(c_{1}^{\ast },\ldots ,c_{d}^{\ast })$ minimizes
the squared prediction error 
\begin{equation}
\mathrm{SPE}(\mathbf{c})=E\{\Vert \Lambda (\cdot ,\mathbf{s}%
_{0})-\sum_{j=1}^{d}c_{j}\Lambda (\cdot ,\mathbf{s}_{j})\Vert ^{2}\}
\label{eq:SPE}
\end{equation}%
subject to the unbiasedness constraint 
\begin{equation}
\mu (t,\mathbf{s}_{0})=\sum_{j=1}^{d}c_{j}\mu (t,\mathbf{s}_{j}),
\label{eq:constraint}
\end{equation}%
where $\mu (t,\mathbf{s})=E\{\Lambda (t,\mathbf{s})\}$ and $\left\Vert \cdot
\right\Vert $ is the $L_{2}$ norm.

The squared prediction error (\ref{eq:SPE}), in view of the constraints (\ref%
{eq:constraint}), comes down to 
\begin{equation}
\mathrm{SPE}(\mathbf{c})=\mathbf{c}^{T}\mathbf{\Sigma c}-2\mathbf{c}^{T}%
\mathbf{\sigma }_{0}+\sigma _{00},  \label{eq:SPE_2}
\end{equation}%
where $\mathbf{\Sigma }$ has elements 
\begin{equation}
\Sigma _{jk}=\int \limfunc{cov}\left\{ \Lambda (t,\mathbf{s}_{j}),\Lambda (t,%
\mathbf{s}_{k})\right\} \ dt,  \label{eq:Sigma_jk}
\end{equation}%
$\mathbf{\sigma }_{0}$ has elements 
\begin{equation}
\sigma _{0j}=\int \limfunc{cov}\left\{ \Lambda (t,\mathbf{s}_{j}),\Lambda (t,%
\mathbf{s}_{0})\right\} \ dt,  \label{eq:sigma_j0}
\end{equation}%
and $\sigma _{00}=\int \func{var}\left\{ \Lambda (t,\mathbf{s}_{0})\right\}
dt$. By multiplying both sides of (\ref{eq:constraint}) by the $\mu (t,%
\mathbf{s}_{k})$'s and integrating $t$ out, the constraints can be expressed
as 
\begin{equation}
\mathbf{Mc}=\mathbf{m}_{0},  \label{eq:constraint_3}
\end{equation}%
where $\mathbf{M}$ has elements 
\begin{equation}
M_{jk}=\int \mu (t,\mathbf{s}_{j})\mu (t,\mathbf{s}_{k})~dt  \label{eq:M_jk}
\end{equation}
and $\mathbf{m}_{0}$ has elements 
\begin{equation}
m_{0j}=\int \mu (t,\mathbf{s}_{j})\mu (t,\mathbf{s}_{0})\ dt.
\label{eq:m_0j}
\end{equation}
When $\mathbf{M}$ is full rank, the minimization of (\ref{eq:SPE_2}) subject
to (\ref{eq:constraint_3})\ has a closed-form solution 
\begin{equation*}
\left[ 
\begin{array}{l}
\mathbf{c}^{\ast } \\ 
\mathbf{\ell }%
\end{array}%
\right] =\left[ 
\begin{array}{ll}
\mathbf{\Sigma } & \mathbf{M}^{T} \\ 
\mathbf{M} & \mathbf{O}%
\end{array}%
\right] ^{-1}\left[ 
\begin{array}{l}
\mathbf{\sigma }_{0} \\ 
\mathbf{m}_{0}%
\end{array}%
\right] ,
\end{equation*}%
where $\mathbf{\ell }\in \mathbb{R}^{d}$ is the Lagrange multiplier and $%
\mathbf{O}$ is the $d\times d$ zero matrix. To compute $\mathbf{c}^{\ast }$,
then, it is necessary to obtain estimators of $\mathbf{\Sigma }$, $\mathbf{%
\sigma }_{0}$, $\mathbf{M}$, and $\mathbf{m}_{0}$, which will be introduced
in Section \ref{sec:Estimation}. 

The $d\times d$ matrix $\mathbf{M}$ is often not full rank. For example, if $%
\mu (t,\mathbf{s}_{j})\equiv \mu (t)$ for all $\mathbf{s}_{j}$, then $%
\mathbf{M}$ has rank one. Even when it has full rank, $\mathbf{M}$ is often
ill-conditioned, with many eigenvalues near zero. In those situations $%
\mathbf{M}$ can be truncated as follows. Let $\mathbf{M}=\mathbf{U\Delta U}%
^{T}$ be the spectral decomposition of $\mathbf{M}$, where $\mathbf{\Delta }=%
\func{diag}(\delta _{1},\ldots ,\delta _{d})$ are the eigenvalues of $%
\mathbf{M}$ in decreasing order and $\mathbf{U}$ is an orthogonal matrix of
eigenvectors. Let $r$ be the number of $\delta _{j}$'s that are strictly
positive; or, in practice, we can take the smallest $r$ such that $%
\sum_{j=1}^{r}\delta _{j}/\sum_{j=1}^{d}\delta _{j}\geq 0.9$, say. Then the $%
d$-dimensional constraints (\ref{eq:constraint_3}) are replaced by the $r$%
-dimensional approximation 
\begin{equation}
\mathbf{\tilde{M}c}=\mathbf{\tilde{m}}_{0},  \label{eq:constraint_tr}
\end{equation}%
where $\mathbf{\tilde{M}}=\mathbf{\Delta }_{r}\mathbf{U}_{r}^{T}$, $\mathbf{%
\tilde{m}}_{0}=\mathbf{U}_{r}^{T}\mathbf{m}_{0}$, $\mathbf{\Delta }_{r}=%
\func{diag}(\delta _{1},\ldots ,\delta _{r})$, and $\mathbf{U}_{r}$ are the
first $r$ columns of $\mathbf{U}$.

We note that although the kriging prediction problem was framed in terms of
the unobservable intensity processes $\Lambda ^{j}(t)$'s, the estimated
kriging coefficients $\mathbf{\hat{c}}^{\ast }$, once obtained, can also be
used for direct real-time prediction of the count functions: for a given day 
$i$, the predicted count function at the new site $\mathbf{s}_{0}$ would be $%
N_{i}^{0\ast }(t)=\sum_{j=1}^{d}c_{j}^{\ast }N_{i}^{j}(t)$, where the $%
N_{i}^{j}(t)$'s are the observed count functions at the sites $\mathbf{s}_{j}
$'s. This is an unbiased predictor, in view of (\ref{eq:constraint}), since $%
E\{N^{0}(t)\}=\int_{a}^{t}\mu (t,\mathbf{s}_{0})dt$ and $E\{N^{j}(t)\}=%
\int_{a}^{t}\mu (t,\mathbf{s}_{j})dt$.

\subsection{Mean and covariance estimation\label{sec:Estimation}}

\subsubsection{Nonparametric estimators at observed sites\label%
{sec:Nonpar_estim}}

Estimation of the mean functions $\mu _{j}(t)=\mu (t,\mathbf{s}_{j})$ and
the covariance functions $\rho _{jk}(t,t^{\prime })=\limfunc{cov}\left\{
\Lambda (t,\mathbf{s}_{j}),\Lambda (t^{\prime },\mathbf{s}_{k})\right\} $ at
the spatial points $\mathbf{s}_{j}$ and $\mathbf{s}_{k}$ where data is
available can be done as follows. Since $X^{j}\mid \Lambda ^{j}=\lambda ^{j}$
is a Poisson process with intensity function $\lambda ^{j}(t)$, for any
integrable functions $f(t)$ and $g(t)$ we have, as shown in the
Supplementary Material, 
\begin{equation}
E\left\{ \sum_{t\in X^{j}}f(t)\right\} =\int f(t)\mu _{j}(t)~dt,
\label{eq:C1}
\end{equation}%
\begin{equation}
E\left\{ \sum_{t\in X^{j}}\sum_{t^{\prime }\in X^{k}}f(t)g(t^{\prime
})\right\} =\iint f(t)g(t^{\prime })R_{jk}(t,t^{\prime })~dt~dt^{\prime }%
\text{, \ for }j\neq k\text{,}  \label{eq:C2}
\end{equation}%
and%
\begin{equation}
E\left\{ \sum_{t\in X^{j}}\sum_{t^{\prime }\in X^{j},t^{\prime }\neq
t}f(t)g(t^{\prime })\right\} =\iint f(t)g(t^{\prime })R_{jj}(t,t^{\prime
})~dt~dt^{\prime },  \label{eq:C3}
\end{equation}%
where $R_{jk}(t,t^{\prime })=E\left\{ \Lambda (t,\mathbf{s}_{j})\Lambda
(t^{\prime },\mathbf{s}_{k})\right\} $. Now consider a $B$-spline basis (De
Boor, 2001, ch.~9) $\mathbf{\beta }(t)=(\beta _{1}(t),\ldots ,\beta
_{p}(t))^{T}$ on $[a,b]$, and let $\mathbf{G}=\int \mathbf{\beta }(t)\mathbf{%
\beta }(t)^{T}~dt$. Then, given independent and identically distributed
replications $\mathbf{X}_{1},\ldots ,\mathbf{X}_{n}$ of the multivariate
process $\mathbf{X}$, define 
\begin{equation}
\hat{\mu}_{j}(t)=\mathbf{\beta }(t)^{T}\mathbf{G}^{-1}\frac{1}{n}%
\sum_{i=1}^{n}\sum_{u\in X_{i}^{j}}\mathbf{\beta }(u),  \label{eq:mu_hat}
\end{equation}%
\begin{equation}
\hat{R}_{jk}(t,t^{\prime })=\mathbf{\beta }(t)^{T}\mathbf{G}^{-1}\left\{ 
\frac{1}{n}\sum_{i=1}^{n}\sum_{u\in X_{i}^{j}}\sum_{v\in X_{i}^{k}}\mathbf{%
\beta }(u)\mathbf{\beta }(v)^{T}\right\} \mathbf{G}^{-1}\mathbf{\beta }%
(t^{\prime })\text{, for }j\neq k\text{,}  \label{eq:R_hat_jk}
\end{equation}%
and 
\begin{equation}
\hat{R}_{jj}(t,t^{\prime })=\mathbf{\beta }(t)^{T}\mathbf{G}^{-1}\left\{ 
\frac{1}{n}\sum_{i=1}^{n}\sum_{u\in X_{i}^{j}}\sum_{v\in X_{i}^{j},v\neq u}%
\mathbf{\beta }(u)\mathbf{\beta }(v)^{T}\right\} \mathbf{G}^{-1}\mathbf{%
\beta }(t^{\prime }).  \label{eq:R_hat_jj}
\end{equation}%
The consistency of these estimators as the number of replications $n$ goes
to infinity is proved in Section \ref{sec:Asymptotics}, and confirmed by
simulations in Section \ref{sec:Simulations}.

From the above $\hat{\mu}_{j}(t)$'s and $\hat{R}_{jk}(t,t^{\prime })$'s we
obtain $\hat{\rho}_{jk}(t,t^{\prime })=\hat{R}_{jk}(t,t^{\prime })-\hat{\mu}%
_{j}(t)\hat{\mu}_{k}(t^{\prime })$. These are plugged into equations (\ref%
{eq:M_jk}) to obtain $\mathbf{\hat{M}}$ and (\ref{eq:Sigma_jk}) to obtain $%
\mathbf{\hat{\Sigma}}$.

It is often the case that $\mathbf{\hat{\Sigma}}$, although full rank, is
ill-conditioned. We found out in our simulation studies that truncating $%
\mathbf{\hat{\Sigma}}$ improves kriging accuracy. As before, let $\mathbf{%
\hat{\Sigma}}=\mathbf{VHV}^{T}$ be the spectral decomposition of $\mathbf{%
\hat{\Sigma}}$, where $\mathbf{H}=\func{diag}(\eta _{1},\ldots ,\eta _{d})$
are the eigenvalues of $\mathbf{\hat{\Sigma}}$ in decreasing order and $%
\mathbf{V}$ is an orthogonal matrix of eigenvectors. Take the smallest $s$
such that $\sum_{j=1}^{s}\eta _{j}/\sum_{j=1}^{d}\eta _{j}\geq 0.9$. Let $%
\mathbf{H}_{s}=\func{diag}(\eta _{1},\ldots ,\eta _{s})$ and let $\mathbf{V}%
_{s}$ be the first $s$ columns of $\mathbf{V}$. Then solve 
\begin{equation*}
\left[ 
\begin{array}{l}
\mathbf{\hat{c}}_{s} \\ 
\mathbf{\hat{\ell}}%
\end{array}%
\right] =\left[ 
\begin{array}{cc}
\mathbf{H}_{s} & \mathbf{V}_{s}^{T}\mathbf{\hat{M}}^{T} \\ 
\mathbf{\hat{M}V}_{s} & \mathbf{O}%
\end{array}%
\right] ^{-1}\left[ 
\begin{array}{l}
\mathbf{V}_{s}^{T}\mathbf{\hat{\sigma}}_{0} \\ 
\mathbf{\hat{m}}_{0}%
\end{array}%
\right] 
\end{equation*}%
and take $\mathbf{\hat{c}}^{\ast }=\mathbf{V}_{s}\mathbf{\hat{c}}_{s}$. Of
course, if $\mathbf{\hat{M}}$ and $\mathbf{\hat{m}}_{0}$ are truncated as in
(\ref{eq:constraint_tr}), then their truncated versions should be used
instead.

\subsubsection{Estimators at the new site\label{sec:Estim_new_sites}}

To estimate $\mathbf{\sigma }_{0}$ and $\mathbf{m}_{0}$, the above mean and
covariance estimators are extended by smoothing to spatial points $\mathbf{s}%
_{0}$ where no data is available. Consider first the mean function $\mu (t,%
\mathbf{s})$. We model this function as $\mathbf{\beta }(t)^{T}\mathbf{%
B\gamma }(\mathbf{s})$, where $\mathbf{\gamma }(\mathbf{s})=(\mathbf{\gamma }%
_{1}(\mathbf{s}),\ldots ,\mathbf{\gamma }_{q}(\mathbf{s}))^{T}$ is a spatial
basis on $R$, a region of $\mathbb{R}^{2}$ that includes $\mathbf{s}_{0}$
and the $\mathbf{s}_{k}$'s, and $\mathbf{B}$ is a matrix of coefficients to
be estimated from the data. In this paper we use tensor-product splines as $%
\mathbf{\gamma }(\mathbf{s})$ (De Boor, 2001, ch.~17), but other
alternatives are possible, like thin-plate splines (Wahba, 1990) or radial
basis functions (Buhmann, 2003).

To estimate $\mathbf{B}$, note that $\hat{\mu}_{j}(t)$ in (\ref{eq:mu_hat})
has the form $\hat{\mu}_{j}(t)=\mathbf{\beta }(t)^{T}\mathbf{\hat{a}}_{j}$,
with 
\begin{equation*}
\mathbf{\hat{a}}_{j}=\mathbf{G}^{-1}\frac{1}{n}\sum_{i=1}^{n}\sum_{u\in
X_{i}^{j}}\mathbf{\beta }(u),
\end{equation*}%
so the penalized least squares estimator of $\mathbf{B}$ is 
\begin{equation*}
\mathbf{\hat{B}}=\arg \min_{\mathbf{B}}\sum_{j=1}^{d}\left\Vert \mathbf{\hat{%
a}}_{j}-\mathbf{B\gamma }(\mathbf{s}_{j})\right\Vert ^{2}+\xi _{B}P_{1}(%
\mathbf{B}),
\end{equation*}%
where $P_{1}(\mathbf{B})$ is a roughness penalty function and $\xi _{B}$ a
smoothing parameter. As explained in the Supplementary Material, if the
roughness of a bivariate function $f(s^{1},s^{2})$ is measured by $\iint
(\sum_{1\leq i,j\leq 2}f_{ij}^{2})ds^{1}ds^{2}$, where $f_{ij}=\partial
^{2}f/\partial s^{i}\partial s^{j}$, then $P_{1}(\mathbf{B})=\func{tr}\left( 
\mathbf{B}^{T}\mathbf{BJ}\right) $, where $\mathbf{J}$ is a matrix that
depends only on $\mathbf{\gamma }(\mathbf{s})$, and the closed form of $%
\mathbf{\hat{B}}$ is 
\begin{equation}
\mathbf{\hat{B}}=\mathbf{A\Gamma }(\mathbf{\Gamma }^{T}\mathbf{\Gamma }+\xi
_{B}\mathbf{J})^{-1},  \label{eq:B_hat}
\end{equation}%
where $\mathbf{\Gamma }=[\mathbf{\gamma }(\mathbf{s}_{1}),\ldots ,\mathbf{%
\gamma }(\mathbf{s}_{d})]^{T}$ and $\mathbf{A}=[\mathbf{\hat{a}}_{1},\ldots ,%
\mathbf{\hat{a}}_{d}]$. Once $\mathbf{\hat{B}}$ is obtained, $\mu (t,\mathbf{%
s}_{0})$ is estimated by $\hat{\mu}(t,\mathbf{s}_{0})=\mathbf{\beta }(t)^{T}%
\mathbf{\hat{B}\gamma }(\mathbf{s}_{0})$ and plugged into (\ref{eq:m_0j}) to
obtain $\mathbf{\hat{m}}_{0}$.

The optimal smoothing parameter $\xi _{B}$ can be chosen by cross-validation
(Hastie et al., 2009, ch.~7). The leave-one-site-out cross-validation
statistic would be 
\begin{eqnarray*}
\mathrm{CV}(\xi _{B}) &=&\frac{1}{d}\sum_{j=1}^{d}\Vert \mathbf{\hat{a}}_{j}-%
\mathbf{\hat{B}}_{(j)}\mathbf{\gamma }(\mathbf{s}_{j})\Vert ^{2} \\
&=&\frac{1}{d}\sum_{j=1}^{d}\frac{\Vert \mathbf{\hat{a}}_{j}-\mathbf{\hat{B}%
\gamma }(\mathbf{s}_{j})\Vert ^{2}}{(1-h_{B,jj})^{2}},
\end{eqnarray*}%
where $\mathbf{\hat{B}}_{(j)}$ is the $\mathbf{s}_{j}$-deleted version of $%
\mathbf{\hat{B}}$, and $h_{B,jj}$ is the $j$th diagonal element of the hat
matrix $\mathbf{H}_{B}=\mathbf{\Gamma }(\mathbf{\Gamma }^{T}\mathbf{\Gamma }%
+\xi _{B}\mathbf{J})^{-1}\mathbf{\Gamma }^{T}$. If $\mathrm{df}_{B}=\func{tr}%
(\mathbf{H}_{B})$ are the degrees of freedom, then $h_{B,jj}\approx \mathrm{%
df}_{B}/d$ and the generalized cross-validation statistic is 
\begin{equation*}
\mathrm{GCV}(\xi _{B})=\frac{1}{d}\sum_{j=1}^{d}\frac{\Vert \mathbf{\hat{a}}%
_{j}-\mathbf{\hat{B}\gamma }(\mathbf{s}_{j})\Vert ^{2}}{(1-\mathrm{df}%
_{B}/d)^{2}}.
\end{equation*}%
The optimal $\hat{\xi}_{B}$ is chosen as the minimizer of $\mathrm{GCV}(\xi
_{B})$.

To estimate $\mathbf{\sigma }_{0}$ we also use spatial smoothing and model $%
\Sigma (\mathbf{s},\mathbf{s}^{\prime })=\int \limfunc{cov}\left\{ \Lambda
(t,\mathbf{s}),\Lambda (t,\mathbf{s}^{\prime })\right\} dt$ by $\mathbf{%
\gamma }(\mathbf{s})^{T}\mathbf{C\gamma }(\mathbf{s}^{\prime })$, with $%
\mathbf{C}$ symmetric. The penalized least squares estimator of $\mathbf{C}$
is 
\begin{equation}
\mathbf{\hat{C}}=\arg \min_{\mathbf{C}}\sum_{j=1}^{d}\sum_{\substack{ k=1 \\ %
k\neq j}}^{d}\left\{ \hat{\Sigma}_{jk}-\mathbf{\gamma }(\mathbf{s}_{j})^{T}%
\mathbf{C\gamma }(\mathbf{s}_{k})\right\} ^{2}+\xi _{C}P_{2}(\mathbf{C}),
\label{eq:C_hat}
\end{equation}%
where, as before, $P_{2}(\mathbf{C})$ is a roughness penalty function and $%
\xi _{C}$ is a smoothing parameter. Here we use only the off-diagonal
elements $\hat{\Sigma}_{jk}$ with $j\neq k$ for estimation, because, in most
applications, the intrinsic variability at each spatial site creates a ridge
that makes $\Sigma (\mathbf{s},\mathbf{s}^{\prime })$ discontinuous at the
diagonal $\mathbf{s}=\mathbf{s}^{\prime }$.

If the roughness of a function $f(s^{1},s^{2},s^{3},s^{4})$ is measured by $%
\iint (\sum_{1\leq i,j,k,l\leq 2}f_{ijkl}^{2})ds^{1}ds^{2}ds^{3}ds^{4}$,
then $P_{2}(\mathbf{C})=\func{tr}\{(\mathbf{CJ})^{2}\}$ with the same $%
\mathbf{J}$ as before. As shown in the Supplementary Material, the closed
form for $\func{vec}(\mathbf{\hat{C}})$ is 
\begin{equation}
\func{vec}(\mathbf{\hat{C}})=\mathbf{\Omega }^{-1}(\mathbf{\Gamma }%
^{T}\otimes \mathbf{\Gamma }^{T})\func{vec}(\mathbf{\hat{\Sigma}}-\func{diag}%
\mathbf{\hat{\Sigma}}),  \label{eq:vec_C_hat}
\end{equation}%
where $\mathbf{\Omega }=\left\{ (\mathbf{\Gamma }^{T}\otimes \mathbf{\Gamma }%
^{T})(\mathbf{I}-\mathbf{E}^{T}\mathbf{E})(\mathbf{\Gamma }\otimes \mathbf{%
\Gamma })+\xi _{C}(\mathbf{J}\otimes \mathbf{J})\right\} $, $\mathbf{E}^{T}%
\mathbf{E}=\sum_{j=1}^{d}\mathbf{e}_{j}\mathbf{e}_{j}^{T}\otimes \mathbf{e}%
_{j}\mathbf{e}_{j}^{T}$ and $\mathbf{e}_{j}$ is the $j$-th canonical vector
in $\mathbb{R}^{d}$. Once $\mathbf{\hat{C}}$ has been obtained, the $\sigma
_{0j}$'s in (\ref{eq:sigma_j0}) are estimated by $\hat{\sigma}_{0j}=\mathbf{%
\gamma }(\mathbf{s}_{j})^{T}\mathbf{\hat{C}\gamma }(\mathbf{s}_{0})$.
Details of numerical implementation are also discussed in the Supplementary
Material, since the large dimension of $\mathbf{\Omega }$ would make
straightforward implementation of (\ref{eq:vec_C_hat}) very inefficient and
time consuming.

The optimal smoothing parameter $\xi _{C}$ can be found, as before, by
generalized cross-validation. The leave-one-$(j,k)$-out cross-validation
statistic is 
\begin{eqnarray*}
\mathrm{CV}(\xi _{C}) &=&\frac{1}{d(d-1)}\sum_{j=1}^{d}\sum_{\substack{ k=1
\\ k\neq j}}^{d}\left\{ \hat{\Sigma}_{jk}-\mathbf{\gamma }(\mathbf{s}%
_{j})^{T}\mathbf{\hat{C}}_{(j,k)}\mathbf{\gamma }(\mathbf{s}_{k})\right\}
^{2} \\
&=&\frac{1}{d(d-1)}\sum_{j=1}^{d}\sum_{\substack{ k=1 \\ k\neq j}}^{d}\frac{%
\left\{ \hat{\Sigma}_{jk}-\mathbf{\gamma }(\mathbf{s}_{j})^{T}\mathbf{\hat{C}%
\gamma }(\mathbf{s}_{k})\right\} ^{2}}{(1-h_{C,(j,k)})^{2}},
\end{eqnarray*}%
where $\mathbf{\hat{C}}_{(j,k)}$ is the $(j,k)$-deleted version of $\mathbf{%
\hat{C}}$ and $h_{C,(j,k)}$ is the diagonal element of the hat matrix $%
\mathbf{H}_{C}=(\mathbf{\Gamma }\otimes \mathbf{\Gamma })\mathbf{\Omega }%
^{-1}(\mathbf{\Gamma }^{T}\otimes \mathbf{\Gamma }^{T})$ corresponding to
the location of $\hat{\Sigma}_{jk}$ in $\func{vec}(\mathbf{\hat{\Sigma})}$.
If $\mathrm{df}_{C}=\func{tr}(\mathbf{H}_{C})$, then $h_{C,(j,k)}\approx 
\mathrm{df}_{C}/d(d-1)$ and the generalized cross-validation statistic is 
\begin{equation*}
\mathrm{GCV}(\xi _{C})=\frac{1}{d(d-1)}\sum_{j=1}^{d}\sum_{\substack{ k=1 \\ %
k\neq j}}^{d}\frac{\left\{ \hat{\Sigma}_{jk}-\mathbf{\gamma }(\mathbf{s}%
_{j})^{T}\mathbf{\hat{C}\gamma }(\mathbf{s}_{k})\right\} ^{2}}{\{1-\mathrm{df%
}_{C}/d(d-1)\}^{2}}.
\end{equation*}%
The optimal $\hat{\xi}_{C}$ is chosen as the minimizer of $\mathrm{GCV}(\xi
_{C})$.

\section{Asymptotics\label{sec:Asymptotics}}

In this section we establish the consistency of the nonparametric estimators
of the mean and covariance functions introduced in Section \ref%
{sec:Nonpar_estim}. The convergence rates that we obtain are in line with
the standard asymptotic results for regression splines (Agarwal and Studden,
1980; Zhou et al., 1998).

We assume that the $B$-spline basis $\mathbf{\beta }(t)$ has order $r$ and
is defined by a knot sequence $\{\tau _{1},\ldots ,\tau _{k}\}$ that is
regular, in the sense that 
\begin{equation*}
\int_{a}^{\tau _{i}}g(t)~dt=\frac{i}{k+1},\ \ i=1,\ldots ,k,
\end{equation*}%
for a strictly positive density function $g(t)$ on $[a,b]$. The basis
dimension $p$ is then $r+k$. The observed point processes $\mathbf{X}%
_{1},\ldots ,\mathbf{X}_{n}$ are assumed to be independent and identically
distributed replications of a $d$-variate doubly stochastic Poisson process $%
\mathbf{X}$ with latent intensity process $\mathbf{\Lambda }$, as explained
in Section \ref{sec:Point_processes}. The norm $\left\Vert \cdot \right\Vert 
$ below is the standard $L_{2}[a,b]$ norm, and $L_{2}^{r}[a,b]$ is the
Sobolev space of functions $f$ such that $D^{r-1}f$ is absolutely continuous
on $[a,b]$ and $D^{r}f\in L_{2}[a,b]$, where $D$ denotes differentiation.
Proofs of the results in this section are given in the Supplementary
Material.

\begin{theorem}
\label{thm:Mu_asymp}Let $\hat{\mu}_{j}(t)$ be the estimator defined in (\ref%
{eq:mu_hat}), and suppose that $\mu _{j}\in L_{2}^{r}[a,b]$. Then 
\begin{equation*}
E\Vert \hat{\mu}_{j}-\mu _{j}\Vert ^{2}=\frac{1}{n}O(k)+O\left( \frac{1}{%
k^{2r}}\right) .
\end{equation*}%
The fastest convergence rate is attained for $k=O\left( n^{1/(2r+1)}\right) $%
, in which case $E\Vert \hat{\mu}_{j}-\mu _{j}\Vert ^{2}=O\left(
n^{-2r/(2r+1)}\right) $.
\end{theorem}

Theorem \ref{thm:Mu_asymp} shows that the optimal nonparametric convergence
rate $O\left( n^{-2r/(2r+1)}\right) $ for functions in $L_{2}^{r}[a,b]$
(Stone, 1982) is attained by $\hat{\mu}_{j}(t)$ if $k$ is chosen
appropriately. For cubic splines, $r=4$ and then the optimal rates are $%
k=O\left( n^{1/9}\right) $ and $E\Vert \hat{\mu}_{j}-\mu _{j}\Vert
^{2}=O\left( n^{-8/9}\right) $. The number of knots $k$, then, should grow
slowly with $n$, since, for example, $400^{1/9}\approx 2$.

The next theorem gives convergence rates for the $\hat{R}_{jk}(t,t^{\prime })
$'s. Now $\left\Vert \cdot \right\Vert $ is the $L_{2}([a,b]\times \lbrack
a,b])$ norm and $L_{2}^{(r,r)}([a,b]\times \lbrack a,b])$ is the tensor
Sobolev space of bivariate functions $f$ such that $D_{i}^{r-1}f$ is
absolutely continuous on $[a,b]\times \lbrack a,b]$ and $D_{i}^{r}f\in
L_{2}([a,b]\times \lbrack a,b])$, where $D_{i}$ denotes differentiation with
respect to the $i$-th variable.

\begin{theorem}
\label{thm:R_asymp}Let $\hat{R}_{jk}(t,t^{\prime })$ be the estimator
defined in (\ref{eq:R_hat_jk}), if $j\neq k$, or in (\ref{eq:R_hat_jj}), if $%
j=k$. Then, if $R_{jk}\in L_{2}^{(r,r)}([a,b]\times \lbrack a,b])$, we have 
\begin{equation*}
E\Vert \hat{R}_{jk}-R_{jk}\Vert ^{2}=\frac{1}{n}O(k^{2})+O\left( \frac{1}{%
k^{2r}}\right) .
\end{equation*}%
The fastest convergence rate is attained for $k=O\left( n^{1/(2r+2)}\right) $%
, in which case $E\Vert \hat{R}_{jk}-R_{jk}\Vert ^{2}=O\left(
n^{-2r/(2r+2)}\right) $.
\end{theorem}

Once again, Theorem \ref{thm:R_asymp} shows that the optimal convergence
rate $O\left( n^{-2r/(2r+2)}\right) $ for bivariate functions (Stone, 1994)
is attained by $\hat{R}_{jk}(t,t^{\prime })$ if $k$ is suitably chosen. For
cubic splines, the optimal $k$ would have rate $O(n^{1/10})$ and the squared
estimation error would have rate $O(n^{-8/10})$. According to Theorems \ref%
{thm:Mu_asymp} and \ref{thm:R_asymp}, the optimal rates for $k$ are
different for the $\hat{\mu}_{j}(t)$'s and the $\hat{R}_{jk}(t,t^{\prime })$%
's. However, for simplicity we use the same spline basis $\mathbf{\beta }(t)$
in all cases.

\section{Simulations\label{sec:Simulations}}

In this section we study by simulation the consistency and convergence rates
of the mean and covariance estimators introduced in Sections \ref%
{sec:Nonpar_estim} and \ref{sec:Estim_new_sites}, and of the kriging
predictors introduced in Section \ref{sec:Kriging}. We are specifically
interested in the effects of sample size $n$, spatial grid size $d$, and
grid spacing $\delta =\min_{j\neq k}\left\Vert \mathbf{s}_{j}-\mathbf{s}%
_{k}\right\Vert $ on estimation and prediction error.

To this end we simulated the following scenarios. Three spatial grids were
considered: \emph{(i)} $d=16$ uniformly spaced $\mathbf{s}_{j}$'s on the
square $[-0.5,0.5]\times \lbrack -0.5,0.5]$, \emph{(ii)} $d=16$ uniformly
spaced $\mathbf{s}_{j}$'s on the square $[-0.2,0.2]\times \lbrack -0.2,0.2]$%
, and \emph{(iii)} $d=64$ uniformly spaced $\mathbf{s}_{j}$'s on the square $%
[-0.5,0.5]\times \lbrack -0.5,0.5]$. The respective grid spacings are \emph{%
(i)} $\delta =0.33$, \emph{(ii)} $\delta =0.13$, and \emph{(iii)} $\delta
=0.14$. Grids \emph{(i)} and \emph{(iii)} cover the same range but a larger $%
d$ makes \emph{(iii)} denser, while grids \emph{(i)} and \emph{(ii)} have
the same size $d$ but \emph{(ii)} is denser because it covers a smaller
range. The kriging predictor was evaluated at the spatial point $\mathbf{s}%
_{0}=(0,0)$.

The latent processes $\Lambda (t,\mathbf{s}_{j})$'s were generated according
to the log-Gaussian model 
\begin{equation}
\Lambda (t,\mathbf{s}_{j})=\exp \{\nu (t)+U_{j}\phi (t)\}
\label{eq:sim_model}
\end{equation}%
for $t\in \lbrack 0,1]$, with $\nu (t)=\sin (\pi t)+\ln 20$ and $\phi (t)=%
\sqrt{2}\sin (\pi t)$. The $U_{j}$'s, which determine the spatial
correlations, were defined as 
\begin{equation}
U_{j}=g(\mathbf{s}_{j})W+E_{j}  \label{eq:sim_U}
\end{equation}%
with $W\sim N(0,0.072)$ and $E_{j}\sim N(0,0.018)$. The $E_{j}$'s were
independent among themselves and of $W$. Two functions $g(\mathbf{s})$ were
considered: Model 1, $g(\mathbf{s})=1/(1+\left\Vert \mathbf{s}\right\Vert )$%
, and Model 2, $g(\mathbf{s})=1$. The common factor $g(\mathbf{s})W$ in (\ref%
{eq:sim_U}) makes $\limfunc{cov}\{\Lambda (t,\mathbf{s}),\Lambda (t,\mathbf{s%
}^{\prime })\}$ a smooth function for $\mathbf{s}\neq \mathbf{s}^{\prime }$,
but the $E_{j}$'s create a ridge at $\mathbf{s}=\mathbf{s}^{\prime }$, as
noted in Section \ref{sec:Estim_new_sites}. Explicit expressions for $\mu (t,%
\mathbf{s}_{j})$ and $\limfunc{cov}\{\Lambda (t,\mathbf{s}_{j}),\Lambda (t,%
\mathbf{s}_{k})\}$ are given in the Supplementary Material.

For estimation we used cubic $B$-splines with five equally-spaced knots as
temporal basis $\mathbf{\beta }(t)$, and tensor-product cubic $B$-splines
with six equally-spaced knots on each coordinate as spatial basis $\mathbf{%
\gamma }(\mathbf{s})$. The respective basis dimensions are $p=9$ and $q=100$%
. The optimal smoothing parameters $\xi _{B}$ and $\xi _{C}$ were chosen by
generalized cross-validation, as explained in Section \ref{sec:Estimation}.
Four sample sizes $n$ were considered: 50, 100, 200, and 400. Each scenario
was replicated 400 times.

We are mainly interested in estimation of the quantities $\mathbf{M}$ and $%
\mathbf{m}_{0}$ in (\ref{eq:constraint_3}), of $\mathbf{\Sigma }$ in (\ref%
{eq:Sigma_jk}),\ and of $\mathbf{\sigma }_{0}$ in (\ref{eq:sigma_j0}),
because they are needed for kriging. For $\mathbf{\hat{M}}$ we define the
relative error measures: $\mathrm{bias}(\mathbf{\hat{M}})=\Vert E\func{vech}%
\mathbf{\hat{M}}-\func{vech}\mathbf{M}\Vert /\left\Vert \func{vech}\mathbf{M}%
\right\Vert $, $\func{sd}(\mathbf{\hat{M}})=\{E\Vert \func{vech}\mathbf{\hat{%
M}}-E\func{vech}\mathbf{\hat{M}}\Vert ^{2}\}^{1/2}/\left\Vert \func{vech}%
\mathbf{M}\right\Vert $, and $\mathrm{rmse}(\mathbf{\hat{M}})=\{E\Vert \func{%
vech}\mathbf{\hat{M}}-\func{vech}\mathbf{M}\Vert ^{2}\}^{1/2}/\left\Vert 
\func{vech}\mathbf{M}\right\Vert $, where $\func{vech}$ denotes the
vectorization of the lower triangular part of a matrix and $\left\Vert \cdot
\right\Vert $ the usual Euclidean norm. Analogous measures are defined for $%
\mathbf{\hat{m}}_{0}$, $\func{vech}\mathbf{\hat{\Sigma}}$, and $\mathbf{\hat{%
\sigma}}_{0}$. To assess the accuracy of the kriging predictor we compared
the best \textrm{SPE} (\ref{eq:SPE}) attained by the true parameters, 
\textrm{SPE}$_{0}$, with the \textrm{SPE} attained by the estimators, $%
\widehat{\mathrm{SPE}}$. Since $\widehat{\mathrm{SPE}}\geq \mathrm{SPE}_{0}$%
, $\mathrm{bias}(\widehat{\mathrm{SPE}})=\mathrm{rmse}(\widehat{\mathrm{SPE}}%
)$. The $\mathrm{rmse}$'s for all parameters are reported in Table \ref%
{tab:Sim_errors}. Biases and standard deviations can be found in the
Supplementary Material.

\begin{table}[tbp] \centering%

\begin{tabular}{llllccccccccccc}
&  &  &  & \multicolumn{5}{c}{Model 1} &  & \multicolumn{5}{c}{Model 2} \\ 
\cline{5-9}\cline{11-15}
\multicolumn{1}{c}{Grid} & \multicolumn{1}{c}{} & \multicolumn{1}{c}{$n$} & 
\multicolumn{1}{c}{} & $\mathbf{M}$ & $\mathbf{m}_{0}$ & $\mathbf{\Sigma }$
& $\mathbf{\sigma }_{0}$ & \textrm{SPE} &  & $\mathbf{M}$ & $\mathbf{m}_{0}$
& $\mathbf{\Sigma }$ & $\mathbf{\sigma }_{0}$ & \textrm{SPE} \\ \hline
&  &  &  &  &  &  &  &  &  & \multicolumn{1}{l}{} & \multicolumn{1}{l}{} & 
\multicolumn{1}{l}{} & \multicolumn{1}{l}{} & \multicolumn{1}{l}{} \\ 
(i) &  & 50 &  & \multicolumn{1}{l}{$.077$} & \multicolumn{1}{l}{$.074$} & 
\multicolumn{1}{l}{$.41$} & \multicolumn{1}{l}{$.43$} & \multicolumn{1}{l}{$%
.80$} &  & \multicolumn{1}{l}{$.110$} & \multicolumn{1}{l}{$.105$} & 
\multicolumn{1}{l}{$.41$} & \multicolumn{1}{l}{$.42$} & \multicolumn{1}{l}{$%
.48$} \\ 
&  & 100 &  & \multicolumn{1}{l}{$.057$} & \multicolumn{1}{l}{$.056$} & 
\multicolumn{1}{l}{$.29$} & \multicolumn{1}{l}{$.40$} & \multicolumn{1}{l}{$%
.51$} &  & \multicolumn{1}{l}{$.070$} & \multicolumn{1}{l}{$.066$} & 
\multicolumn{1}{l}{$.28$} & \multicolumn{1}{l}{$.34$} & \multicolumn{1}{l}{$%
.27$} \\ 
&  & 200 &  & \multicolumn{1}{l}{$.042$} & \multicolumn{1}{l}{$.043$} & 
\multicolumn{1}{l}{$.22$} & \multicolumn{1}{l}{$.34$} & \multicolumn{1}{l}{$%
.51$} &  & \multicolumn{1}{l}{$.048$} & \multicolumn{1}{l}{$.045$} & 
\multicolumn{1}{l}{$.19$} & \multicolumn{1}{l}{$.29$} & \multicolumn{1}{l}{$%
.41$} \\ 
&  & 400 &  & \multicolumn{1}{l}{$.027$} & \multicolumn{1}{l}{$.031$} & 
\multicolumn{1}{l}{$.14$} & \multicolumn{1}{l}{$.35$} & \multicolumn{1}{l}{$%
.61$} &  & \multicolumn{1}{l}{$.041$} & \multicolumn{1}{l}{$.040$} & 
\multicolumn{1}{l}{$.12$} & \multicolumn{1}{l}{$.27$} & \multicolumn{1}{l}{$%
.61$} \\ 
&  &  &  & \multicolumn{1}{l}{} & \multicolumn{1}{l}{} & \multicolumn{1}{l}{}
& \multicolumn{1}{l}{} & \multicolumn{1}{l}{} &  & \multicolumn{1}{l}{} & 
\multicolumn{1}{l}{} & \multicolumn{1}{l}{} & \multicolumn{1}{l}{} & 
\multicolumn{1}{l}{} \\ 
(ii) &  & 50 &  & \multicolumn{1}{l}{$.089$} & \multicolumn{1}{l}{$.080$} & 
\multicolumn{1}{l}{$.35$} & \multicolumn{1}{l}{$.28$} & \multicolumn{1}{l}{$%
.07$} &  & \multicolumn{1}{l}{$.102$} & \multicolumn{1}{l}{$.097$} & 
\multicolumn{1}{l}{$.43$} & \multicolumn{1}{l}{$.39$} & \multicolumn{1}{l}{$%
.05$} \\ 
&  & 100 &  & \multicolumn{1}{l}{$.065$} & \multicolumn{1}{l}{$.060$} & 
\multicolumn{1}{l}{$.27$} & \multicolumn{1}{l}{$.22$} & \multicolumn{1}{l}{$%
.05$} &  & \multicolumn{1}{l}{$.073$} & \multicolumn{1}{l}{$.070$} & 
\multicolumn{1}{l}{$.25$} & \multicolumn{1}{l}{$.23$} & \multicolumn{1}{l}{$%
.03$} \\ 
&  & 200 &  & \multicolumn{1}{l}{$.043$} & \multicolumn{1}{l}{$.041$} & 
\multicolumn{1}{l}{$.18$} & \multicolumn{1}{l}{$.17$} & \multicolumn{1}{l}{$%
.04$} &  & \multicolumn{1}{l}{$.048$} & \multicolumn{1}{l}{$.045$} & 
\multicolumn{1}{l}{$.19$} & \multicolumn{1}{l}{$.17$} & \multicolumn{1}{l}{$%
.02$} \\ 
&  & 400 &  & \multicolumn{1}{l}{$.032$} & \multicolumn{1}{l}{$.030$} & 
\multicolumn{1}{l}{$.13$} & \multicolumn{1}{l}{$.13$} & \multicolumn{1}{l}{$%
.03$} &  & \multicolumn{1}{l}{$.039$} & \multicolumn{1}{l}{$.037$} & 
\multicolumn{1}{l}{$.13$} & \multicolumn{1}{l}{$.12$} & \multicolumn{1}{l}{$%
.02$} \\ 
&  &  &  & \multicolumn{1}{l}{} & \multicolumn{1}{l}{} & \multicolumn{1}{l}{}
& \multicolumn{1}{l}{} & \multicolumn{1}{l}{} &  & \multicolumn{1}{l}{} & 
\multicolumn{1}{l}{} & \multicolumn{1}{l}{} & \multicolumn{1}{l}{} & 
\multicolumn{1}{l}{} \\ 
(iii) &  & 50 &  & \multicolumn{1}{l}{$.076$} & \multicolumn{1}{l}{$.068$} & 
\multicolumn{1}{l}{$.38$} & \multicolumn{1}{l}{$.27$} & \multicolumn{1}{l}{$%
.23$} &  & \multicolumn{1}{l}{$.110$} & \multicolumn{1}{l}{$.105$} & 
\multicolumn{1}{l}{$.38$} & \multicolumn{1}{l}{$.36$} & \multicolumn{1}{l}{$%
.18$} \\ 
&  & 100 &  & \multicolumn{1}{l}{$.054$} & \multicolumn{1}{l}{$.054$} & 
\multicolumn{1}{l}{$.29$} & \multicolumn{1}{l}{$.22$} & \multicolumn{1}{l}{$%
.21$} &  & \multicolumn{1}{l}{$.071$} & \multicolumn{1}{l}{$.067$} & 
\multicolumn{1}{l}{$.26$} & \multicolumn{1}{l}{$.23$} & \multicolumn{1}{l}{$%
.13$} \\ 
&  & 200 &  & \multicolumn{1}{l}{$.036$} & \multicolumn{1}{l}{$.042$} & 
\multicolumn{1}{l}{$.20$} & \multicolumn{1}{l}{$.18$} & \multicolumn{1}{l}{$%
.19$} &  & \multicolumn{1}{l}{$.053$} & \multicolumn{1}{l}{$.050$} & 
\multicolumn{1}{l}{$.18$} & \multicolumn{1}{l}{$.15$} & \multicolumn{1}{l}{$%
.09$} \\ 
&  & 400 &  & \multicolumn{1}{l}{$.027$} & \multicolumn{1}{l}{$.033$} & 
\multicolumn{1}{l}{$.15$} & \multicolumn{1}{l}{$.16$} & \multicolumn{1}{l}{$%
.17$} &  & \multicolumn{1}{l}{$.033$} & \multicolumn{1}{l}{$.031$} & 
\multicolumn{1}{l}{$.13$} & \multicolumn{1}{l}{$.11$} & \multicolumn{1}{l}{$%
.06$}%
\end{tabular}

\caption{Simulation Results. Relative root mean squared errors of parameter
estimators.}\label{tab:Sim_errors}%
\end{table}%

We see in Table \ref{tab:Sim_errors} that the mean and covariance estimators
at the observed sites, $\mathbf{\hat{M}}$ and $\mathbf{\hat{\Sigma}}$, are
consistent as $n$ increases, and the magnitudes of the errors do not depend
on the grid, as expected. The estimation errors of $\mathbf{\hat{m}}_{0}$ do
not depend on the grid either because, for these models, the mean functions $%
\mu (t,\mathbf{s}_{j})$ are nearly identical for all $\mathbf{s}_{j}$'s.
However, the situation is different for the covariance estimator $\mathbf{%
\hat{\sigma}}_{0}$, since the true covariance function does change
substantially with $\mathbf{s}$. The errors are larger for the sparser grid 
\emph{(i)}, and smaller for grids \emph{(ii)} and \emph{(iii)}, being of
comparable size for the last two. The behavior of $\mathbf{\hat{\sigma}}_{0}$%
, then, fundamentally depends on grid spacing, not grid size. The accuracy
of the kriging predictor, on the other hand, is better under grid \emph{(ii)}
than under grid \emph{(iii)}, which shows that under comparable grid
spacings, a smaller and more parsimonious grid is generally preferable.
However, this behavior is model dependent: for Model 2, where $g(\mathbf{s})$
does not decrease away from $\mathbf{s}_{0}$, spatial sites further away
from $\mathbf{s}_{0}$ contribute more to prediction than under Model 1, so
the error magnitudes under grids \emph{(ii)} and \emph{(iii)} are not as
different for Model 2 as for Model 1.

\section{Application: predicting bike demand\label{sec:Example}}

As an example of application, in this section we analyze data from the
bicycle-sharing system of the city of Chicago, known as Divvy. The data is
publicly available at the Chicago Data Portal website,
https://data.cityofchicago.org. We will analyze bike trips that took place
on laborable days of 2016, i.e.~weekdays that were not holidays, in the
downtown area known as `the Loop', which is delimited by avenues Grand,
Roosevelt and Halsted on the north, south and west, respectively, and the
lake front on the east. There were 68 active stations in this area during
this period. We specifically study bike check-out times, which can be
modelled as replicated temporal Poisson processes; there are $n=254$
replications in this sample, corresponding to the laborable days of 2016.

\FRAME{ftbpFU}{5.2892in}{3.845in}{0pt}{\Qcb{Divvy data analysis. (a,b)
Observed and (c,d) predicted daily count functions for (a,c) Union station
and (b,d) Lasalle station.}}{\Qlb{fig:Count_functions}}{counts.eps}{\special%
{language "Scientific Word";type "GRAPHIC";maintain-aspect-ratio
TRUE;display "ICON";valid_file "F";width 5.2892in;height 3.845in;depth
0pt;original-width 7.5161in;original-height 5.4518in;cropleft "0";croptop
"1";cropright "1";cropbottom "0";filename '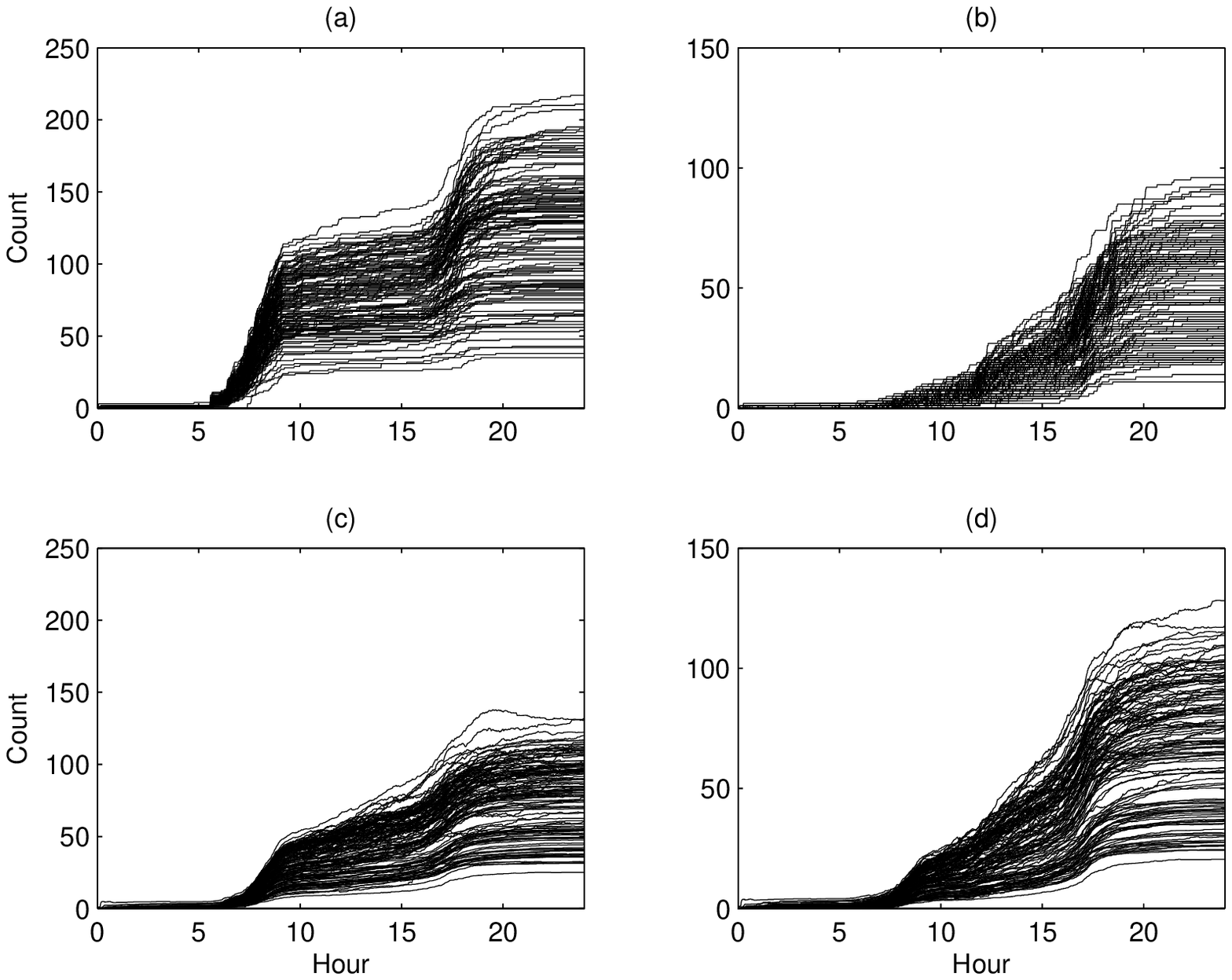';file-properties
"XNPEU";}}

We will set aside two of the 68 bike stations for prediction: the one at the
Union train station, on Adams and Canal Streets, and the one at Lasalle
Avenue and Calhoun Street. These two stations exhibit two very different
usage patterns: the Union station shows a bimodal pattern with peaks at 8am
and 5pm, the morning and evening work commutes, whereas the Lasalle station
shows a unimodal pattern with peak at 5pm, the afternoon work commute. Their
daily count functions are shown in Figure \ref{fig:Count_functions}(a)-(b).
Kriging prediction for these two stations will then be based on the other $%
d=66$ stations in the Loop.

For estimation we used cubic $B$-splines with five equally-spaced knots as
temporal basis $\mathbf{\beta }(t)$ and tensor-product cubic $B$-splines
with six equally-spaced knots on each coordinate as spatial basis $\mathbf{%
\gamma }(\mathbf{s})$. The optimal smoothing parameters $\xi _{B}$ and $\xi
_{C}$ were chosen by generalized cross-validation. Prediction accuracy can
be assessed by comparing the observed count functions $N_{i}(t)$ with the
predicted counts $\hat{N}_{i}(t)$. The root average squared error $%
\{\sum_{i=1}^{n}\Vert N_{i}-\hat{N}_{i}\Vert ^{2}/n\}^{1/2}$, where $%
\left\Vert \cdot \right\Vert $ denotes the $L^{2}$ norm, is $184.2$ for the
Union station and $70.9$ for the Lasalle station.

It is not surprising that the Union station is harder to predict. This bike
station is situated at a train station and therefore has a peculiar pattern
of demand, which is not shared by most other locations in the Loop. In
contrast, the Lasalle station shows a more common pattern of demand. The
average daily count for all the Loop stations is $56.9$, similar to the
average daily count of $52.7$ for the Lasalle station, while the Union
station has a much higher average daily count of $124.4$. This sharp change
in $\mu (t,\mathbf{s})$ at the Union station makes $\mu (t,\mathbf{s}_{0})$
hard to estimate accurately.

\FRAME{ftbpFU}{6.346in}{3.237in}{0pt}{\Qcb{Divvy data analysis. (a,d) Best,
(b,e) median and (c,f) worst fits for (a-c) Union station and (d-f) Lasalle
station. (-----) observed count function, (- - -) predicted count function.}%
}{\Qlb{fig:Best_med_worst}}{best_worst.eps}{\special{language "Scientific
Word";type "GRAPHIC";maintain-aspect-ratio TRUE;display "ICON";valid_file
"F";width 6.346in;height 3.237in;depth 0pt;original-width
9.8874in;original-height 4.5844in;cropleft "0.0870";croptop "1";cropright
"1";cropbottom "0";filename '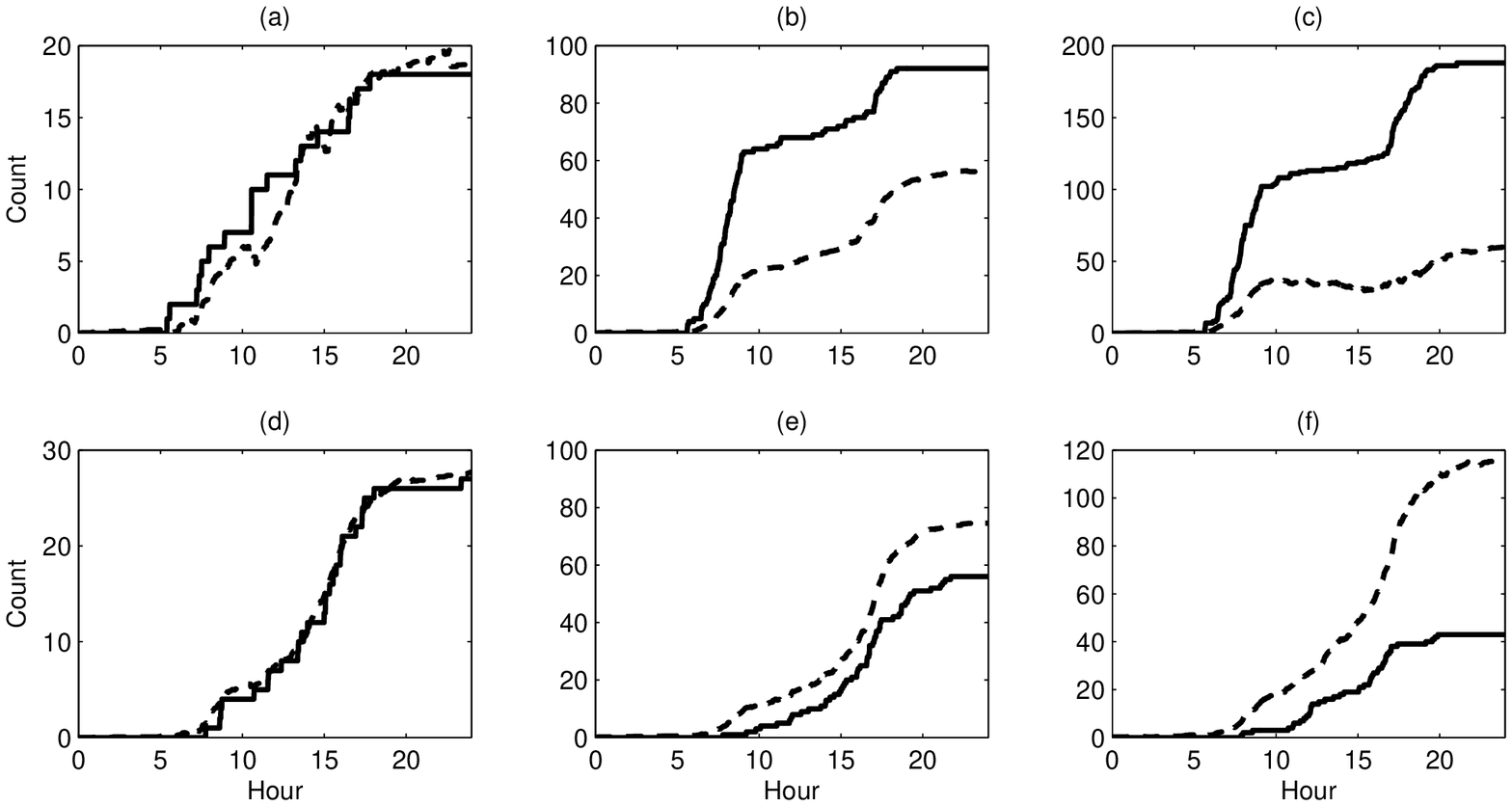';file-properties "XNPEU";}}

The daily predicted counts are shown in Figure \ref{fig:Count_functions}%
(c)-(d). The predictors capture the overall patterns of demand at both
stations, but the morning commute peak is underestimated for the Union
station. Figure \ref{fig:Best_med_worst} shows the best, median, and worst
fits for each station. We see that the kriging predictor tends to
underestimate the counts for the Union station and to overestimate them for
the Lasalle station, but overall, prediction is accurate for the latter.

This example highlights both the possibilities and the limitations of
spatial kriging for these types of problems. Prediction accuracy depends on
the intrinsic variability at each station, which is independent of the other
stations and therefore cannot be predicted, and on the degree of smoothness
of the mean function $\mu (t,\mathbf{s})$ and the covariance functions $%
\Sigma (\mathbf{s}_{j},\mathbf{s})$ at $\mathbf{s}=\mathbf{s}_{0}$. As long
as the intrinsic variability is relatively low and there are no sharp peaks
or troughs in $\mu (t,\mathbf{s})$ and $\Sigma (\mathbf{s}_{j},\mathbf{s})$
at $\mathbf{s}=\mathbf{s}_{0}$, prediction will be accurate. But local
landmarks like train stations, theaters, and stadiums, introduce spatial
discontinuities that make prediction inaccurate when the estimators of $\mu
(t,\mathbf{s}_{0})$ and $\Sigma (\mathbf{s}_{j},\mathbf{s}_{0})$ are only
based on spatial smoothing. Prediction in these situations can likely be
improved by introducing proximity to landmarks as covariates in the model,
but this is a matter for further research.

\section*{References}

\begin{description}
\item Agarwal, G.G., and Studden, W.J. (1980). Asymptotic integrated mean
square error using least squares and bias minimizing splines. \emph{The
Annals of Statistics }\textbf{8} 1307--1325.

\item Buhmann, M.D. (2003). \emph{Radial Basis Functions : Theory and
Implementations}. Cambridge University Press, Cambridge, UK.

\item Cressie, N. (1993). \emph{Statistics for Spatial Data}. John Wiley \&
Sons, New York.

\item De Boor, C. (2001). \emph{A Practical Guide to Splines, Revised Edition%
}. Springer, New York.

\item Gervini, D., and Khanal, M. (2019). Exploring patterns of demand in
bike sharing systems via replicated point process models. \emph{Journal of
the Royal Statistical Society Series C: Applied Statistics} \textbf{68}
585--602.

\item Giraldo, R., Delicado, P., and Mateu, J. (2010). Continuous
time-varying kriging for spatial prediction of functional data: An
environmental application. \emph{Journal of Agricultural, Biological, and
Environmental Statistics} \textbf{15} 66--82.

\item Giraldo, R., Delicado, P., and Mateu, J. (2011). Ordinary kriging for
function-valued spatial data. \emph{Environmental and Ecological Statistics} 
\textbf{18} 411--426.

\item Hastie, T., Tibshirani, R., and Friedman, J. (2009). \emph{The
Elements of Statistical Learning. Data Mining, Inference, and Prediction.
Second Edition. }Springer, New York.

\item Menafoglio, A., Secchi, P., and Dalla Rosa, M. (2013). A universal
kriging predictor for spatially dependent functional data of a Hilbert
space. \emph{Electronic Journal of Statistics} \textbf{7} 2209--2240.

\item M\o ller, J., and Waagepetersen, R.P. (2004). \emph{Statistical
Inference and Simulation for Spatial Point Processes}. Chapman and Hall/CRC,
Boca Raton.

\item Nair, R., and Miller-Hooks, E. (2011). Fleet management for vehicle
sharing operations. \emph{Transportation Science }\textbf{45 }524--540.

\item Shaheen, S., Guzman, S., and Zhang, H. (2010). Bike sharing in Europe,
the Americas and Asia: Past, present and future. \emph{Transportation
Research Record: Journal of the Transportation Research Board }\textbf{2143}
159--167.

\item Stone, C. (1982). Optimal global rates of convergence for
nonparametric regression. \emph{The Annals of Statistics }\textbf{10}
1040--1053.

\item Stone, C. (1994). The use of polynomial splines and their tensor
products in multivariate function estimation. \emph{The Annals of Statistics 
}\textbf{22} 118--184.

\item Wahba, G. (1990). \emph{Spline Models for Observational Data}. Society
for Industrial and Applied Mathematics (SIAM), Philadelphia.

\item Zhou, S., Shen, X., and Wolfe, D.A. (1998). Local asymptotics for
regression splines and confidence region. \emph{The Annals of Statistics }%
\textbf{26} 1760--1782.
\end{description}

\end{document}